# The Stoichiometry of FeSe

A. J. Williams, T. M. McQueen and R. J. Cava

Department of Chemistry, Princeton University, Princeton NJ 08544

**Abstract**

Tetragonal iron selenide, FeSe, the layered parent compound of the recently discovered superconducting arsenide family, has previously been shown to be non magnetic and superconducting with a critical temperature near 8 K. There has, however, been a lack of consensus as to whether selenium vacancies present due to large deviations from ideal stoichiometry are required to give rise to the superconductivity. Here we describe the results of experiments that demonstrate simply that superconducting iron selenide can only be synthesized as a pure material when near stoichiometric (i.e. FeSe). Significant selenium deficiency or excess gives rise to secondary magnetic phases, and a suppression of the superconductivity.

**Introduction**

Since the recent discovery of superconductivity in LnFeAsO$_{1-x}$F$_x$ with superconducting transition temperatures (T$_c$'s) as high as 55K [1,2], a large number of studies have been conducted on the iron pnictide family. Ostensibly the simplest of these materials is tetragonal iron selenide, which consists of layers of edge-sharing FeSe tetrahedra, discovered to be superconducting at 8.5K [3]. This comparative chemical simplicity of FeSe might make it a perfect candidate for study of the interplay of structure, magnetism and superconductivity in this superconducting family. Discrepancies have arisen in the literature regarding its true stoichiometry, however, that suggest that many of the results reported to date may have been obtained on multiple phase materials.

The report describing the discovery of superconductivity in tetragonal FeSe [3] suggested that superconductivity was only observed in samples intentionally prepared with significant selenium deficiency, and that the stoichiometry of the superconducting phase was between $FeSe_{0.82}$ and $FeSe_{0.88}$. The first powder neutron diffraction experiment on this material also concluded that selenium vacancies were present, suggesting a stoichiometry of $FeSe_{0.92}$ [4]. Theoretical treatments of this phase have either assumed the stoichiometry to be FeSe [5,6], without accounting for the critical differences in electron count of 0.16 – 0.36 e/Fe that would be present for selenium-deficient materials, or have specifically addressed the importance of the vacancies in giving rise to the superconductivity [7]. Many physical measurements have been reported on samples that have been prepared with the assumption that iron selenide is deficient in Se [8-12].

On the other hand, several studies have been conducted taking extreme care in the chemical synthesis of iron selenide, particularly with regards to the exclusion of oxygen and water, and these studies have concluded that phase pure materials can only be produced when the starting mixtures are very close to stoichiometric [13-15]. Furthermore, new synchrotron [15,16] and neutron [17,18] powder diffraction data, as well as energy dispersive spectroscopy (EDS) studies [19] have similarly concluded that the stoichiometry of the superconducting phase in iron selenide is at, or close to, 1:1. Nonetheless, work continues to be presented, as described above, with the assumption that the material is highly non-stoichiometric, suggesting that the arguments to the contrary presented so far are not sufficiently transparent to non-chemical researchers in the field to change the prevailing view.

Here we describe a series of experiments designed to demonstrate the true stoichiometry of FeSe in a straightforward fashion. We show that even simple laboratory-based experiments indicate that phase pure samples of iron selenide can only be prepared at nearly stoichiometric compositions, near a perfect 1:1 ratio of Fe to Se, and that deviation from this composition leads directly to the presence of significant impurity phases, observable by both X-ray diffraction and magnetization measurements.

**Experimental**

In total seven samples were synthesized – three substoichiometric samples corresponding to the formulae often quoted in the literature ($FeSe_{0.82}$, $FeSe_{0.87}$ and $FeSe_{0.92}$), one near-stoichiometric sample ($FeSe_{0.99}$), and three samples with selenium excess ($FeSe_{1.04}$, $FeSe_{1.09}$ and $FeSe_{1.14}$). All samples were prepared from iron pieces (Johnson-Matthey, 99.98%) and selenium shot (Alfa-Aesar, 99.999%). The appropriate quantities of freshly polished iron and selenium shot were loaded into cleaned and dried silica tubes. A piece of cleaned carbon, an oxygen getterer, was placed at the opposite end of the tube (and prevented from coming into contact with the sample) and the tube sealed under vacuum. These tubes were then sealed inside a second evacuated silica ampoule, and placed in a furnace at 750 ºC. The temperature was slowly ramped up to 1075 ºC over the course of 4 days, and then held at that temperature for a further 24 hours. This high temperature step is required for scrubbing oxygen from the system and obtaining a homogeneous product. The temperature was then rapidly decreased to 420 ºC, held for an additional 48 hours, and then reduced to 330 ºC for a final annealing step of 2-5 days. Finally, the tubes were quenched into -13 ºC brine, which is required to avoid the low temperature decomposition of the superconducting phase [15]. All samples are stable for short periods of time in air, but were protected from oxidation by storage in an argon glovebox.

The polycrystalline samples obtained were studied by laboratory powder X-ray diffraction (XRD) using a Bruker D8 Focus employing Cu K$\alpha$ radiation and a graphite diffracted beam monochromator. Patterns for all samples were Rietveld analysed using the GSAS software package [20]. Temperature-dependent magnetization and electronic transport properties were measured in a Quantum Design physical property measurement system (PPMS).

**Results and Discussion**

Laboratory XRD data for the selenium deficient and nearly stoichiometric samples are shown in Fig. 1a. A small 15 degree region has been selected to include the strongest reflection from elemental iron, as well as a series of reflections from tetragonal iron selenide. A peak from elemental iron is clearly observed in all the selenium deficient samples, whereas the FeSe$_{0.99}$ sample appears completely clean. A small, characteristic region of the laboratory XRD data for the samples prepared with selenium excess is shown in Fig. 1b. The effects on this side of the phase diagram are even more stark. FeSe$_{1.04}$ is a mixture of the tetragonal form of FeSe, and the hexagonal NiAs-type structure, which has long been known [21] to be non-stoichiometric with a formula Fe$_7$Se$_8$. With increasing selenium content, the proportion of the Fe$_7$Se$_8$ phase increases, until by FeSe$_{1.14}$ (an Fe:Se ratio of approximately 7:8) it is almost single phase. Thus the laboratory x-ray diffraction data show clearly that materials prepared away from the ideal stoichiometry, including those at the often quoted formulas, are multiple phase.

To provide additional evidence for this, magnetization measurements were also performed. Fig 2a shows M(H) curves for the selenium deficient samples, measured at 150 K from 0 to 9 T. All samples show a rapidly saturating magnetization, characteristic of iron metal – however this signal is an order of magnitude higher in the most selenium deficient samples when compared to FeSe$_{0.99}$. The XRD results show that the FeSe$_{0.99}$ sample is dominated by the tetragonal FeSe phase, which has previously been shown to exhibit no long-lived magnetism. These results clearly support the x-ray results that show that large amounts of elemental iron are present in all samples prepared with selenium deficiency. Fig. 2b shows the M(H) curves of the samples made with excess Se, again measured at 150 K between 0 and 9 T. Fe$_7$Se$_8$ is ferrimagnetic at 150 K [22], with a much slower saturating magnetic signal than observed for Fe metal. The magnetic signal in the system can readily be seen to grow as the selenium excess is increased. This is due to the increasing relative proportion of Fe$_7$Se$_8$.

Fig. 3a shows the variation of magnetization at 9 T and 150 K with selenium content in FeSe$_{1\pm\delta}$. It is clear that the minimum occurs very close to a stoichiometric ratio of Fe:Se. This is coupled with diffraction data, which show secondary impurity phases for all samples not prepared at this near-stoichiometric composition. Fig. 3b shows low field magnetization measurements on three samples – one stoichiometric, one with selenium vacancies, and one with a selenium excess. The strongest superconducting signal is clearly observed for FeSe$_{0.99}$, with the non-stoichiometric samples exhibiting both a poorer superconducting transition, and an increased overall magnitude due to the presence of ferromagnetic impurities, in agreement with previous studies [18]. Taken together, the data clearly show that single phase tetragonal iron selenide can only be synthesized at close to 1:1 stoichiometry, with any selenium excess leading to the appearance of secondary phases, consistent with the early phase diagrams [23,24]. They further show that the nearly stoichiometric tetragonal phase is the superconductor. Details of the chemistry of the system very close to the stoichiometric FeSe composition, including the extreme sensitivity of superconductivity to small deviations from stoichiometry, have been reported elsewhere [15].

**Conclusions**

We have presented straightforward evidence that the correct formula for superconducting tetragonal iron selenide is $FeSe_{0.99}$ (or $Fe_{1.01}Se$) and that substantial quantities of selenium vacancies are not a requirement for superconductivity in this material. We submit that future studies of FeSe should reflect this. Samples prepared at the nominal stoichiometry $FeSe_{0.82}$ or $FeSe_{0.92}$ therefore contain 18 or 8 mole percent of their iron as iron metal in circumstances where oxygen has been rigorously excluded from the synthesis. In syntheses where oxygen is not excluded, the samples prepared at these compositions contain 18 or 8 mole percent iron in an $Fe_3O_4$ impurity phase, and some oxygen contamination in FeSe itself [15], with the impact of the oxide impurity on experimental characterization depending in detail on the measurement being performed. Theoretical treatment of this phase should consider the formula to be FeSe. The changes in superconducting properties induced by subtle changes in Fe stoichiometry near the ideal 1:1 ratio in FeSe [15] are a complexity that, much like the 1/8 doping anomaly in cuprate superconductors [25], or the extreme sensitivity of superconductivity to impurities in $Sr_2RuO_4$ [26], requires further consideration.

**Acknowledgements**

The work at Princeton was supported by the US Department of Energy, Division of Basic Energy Sciences, Grant DE-FG02-98ER45706. T. M. McQueen gratefully acknowledges support of the National Science Foundation Graduate Research Fellowship program.

**Figure Captions**

**Figure 1** Selected regions of the laboratory powder X-ray diffraction patterns for (a) $FeSe_{1-\delta}$ and (b) $FeSe_{1+\delta}$

**Figure 2** (a) Variation of magnetization of $FeSe_{1-\delta}$ samples with magnetic field. Ferromagnetic iron impurities are clearly evident in all selenium deficient samples. (b) Variation of magnetization of $FeSe_{1+\delta}$ samples with magnetic field. Ferrimagnetic hexagonal $Fe_7Se_8$ impurities are clearly evident in all samples with selenium excess.

**Figure 3** (a) Variation of 9 T magnetization of $FeSe_{1\pm\delta}$ samples with $\delta$. The minimum clearly occurs close to perfect FeSe stoichiometry (b) Low field magnetization data of various $FeSe_{1\pm\delta}$ samples, showing that the strongest superconducting signal occurs for the most stoichiometric sample.

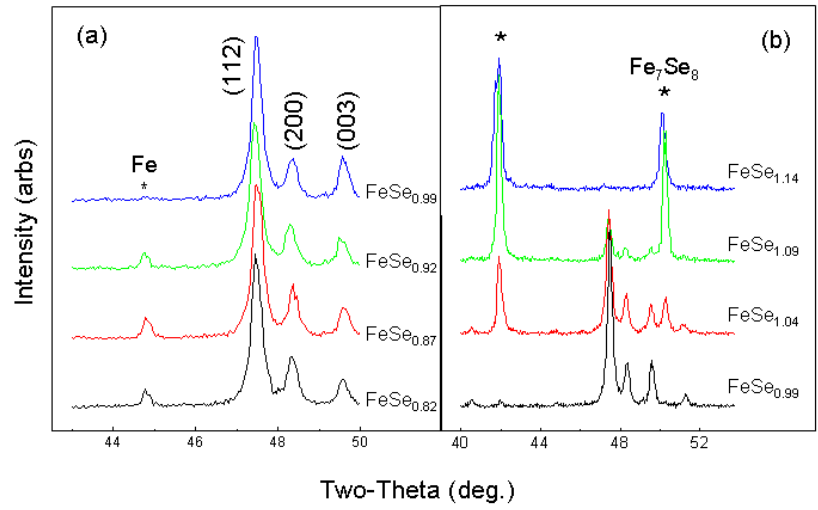

**Fig. 1**

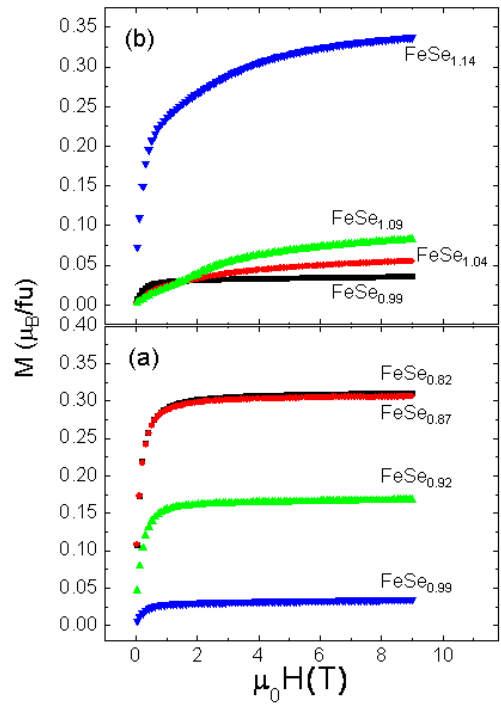

**Fig. 2**

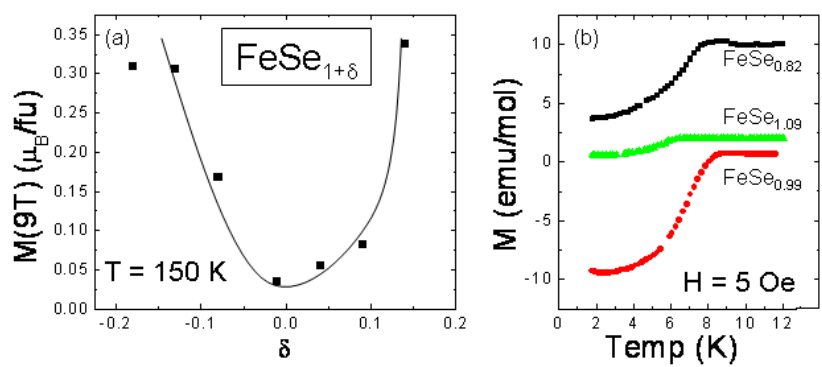

**Fig. 3**